\documentclass[12pt]{article}
\usepackage{graphicx,amsmath,amssymb,euscript,psfrag,epsf}
\usepackage{a4wide}
\newcommand{\be}{\begin{equation}}
\newcommand{\ee}{\end{equation}}
\newcommand{\bea}{\begin{eqnarray}}
\newcommand{\eea}{\end{eqnarray}}
\newcommand{\nl}{\nonumber \\}

\newcommand{\hcbar}{\bar{hc}}
\newcommand{\scbar}{\bar{sc}}
\def\slash#1{#1 \hskip-0.45em /}
\def\Slash#1{#1 \hskip-0.59em /}

\def\beq{\begin{eqnarray}}
\def\eeq{\end{eqnarray}}

\def\eps{\epsilon}

\def\be{\begin{equation}}
\def\ee{\end{equation}}

\def\np{n_+}
\def\nm{n_-}

\begin{document}

\begin{titlepage}

\begin{flushright}
SI-HEP-2005-07 \\
SFB/CPP-05-27 \\
{\tt hep-ph/0506269}\\[0.2cm]
\today
\end{flushright}

\vspace{1.2cm}
\begin{center}
{\Large\bf  Non-factorizable contributions  to deep inelastic scattering 
at large $x$ }
 \end{center}

\vspace{0.5cm}
\begin{center} { Ben D. Pecjak} \\[0.1cm]
{ Theoretische Physik 1, Fachbereich Physik,
Universit\"at Siegen\\ D-57068 Siegen, Germany}
\end{center}

\vspace{0.8cm}
\begin{abstract}
\vspace{0.2cm}\noindent
We use soft-collinear effective theory (SCET) to
study the factorization properties of deep inelastic scattering
in the region of phase space where 
$(1-x)\sim \Lambda_{\rm QCD}/Q$.  By applying a regions analysis to
loop diagrams in the Breit frame, we show that the  
appropriate version of SCET includes anti-hard-collinear,
collinear, and   soft-collinear fields.  We find that
the effects of the soft-collinear fields spoil  
perturbative factorization even at leading order in the $1/Q$ expansion.

\end{abstract}

\end{titlepage}

\section{Introduction}
This paper deals with the factorization properties of deep
inelastic scattering (DIS) in the region of phase space where
$1-x\sim \Lambda_{\rm QCD}/Q$, with $Q$  the large 
energy carried by the virtual photon.  In this kinematical
region the final-state jet carries an 
energy of order $Q$, but has a small invariant mass
$p_x^2=Q^2(1-x)/x\sim Q\Lambda_{\rm QCD}$. 
We assume that perturbation theory is valid at both the 
hard scale $Q^2$ and the jet scale $Q \Lambda_{\rm QCD}$. 
The invariant mass of the target proton defines a third,  
non-perturbative scale $M_P^2 \sim \Lambda_{\rm QCD}^2$.
In similar cases in inclusive $B$ decay it is  possible to
derive  factorization formulas which separate the physics
from the three scales 
$Q^2\gg Q\Lambda_{\rm QCD}\gg\Lambda_{\rm QCD}^2$ into a
convolution of the  generic form \cite{Korchemsky:1994jb}
\be\label{eq:fact}
H \cdot J \otimes S.
\ee
The functions $H$ and $J$ are perturbatively calculable
hard and jet functions depending on fluctuations at
the scales $Q^2$ and $Q\Lambda_{\rm QCD}$ respectively, 
and $S$ is a non-perturbative function 
containing  physics at the low-energy
scale $\Lambda_{\rm QCD}^2$. The symbol $\otimes$ stands
for a convolution. Our goal is to use effective field
theory methods to establish whether such a
factorization formula can be derived for deep inelastic 
scattering in the large-$x$ limit.

Recent studies of perturbative factorization in $B$ decay
have relied heavily on soft-collinear effective
theory (SCET)
\cite{Bauer:2000yr, Bauer:2001yt, Beneke:2002ph, Beneke:2002ni}.  
These include many applications to 
inclusive decay, both at leading order
\cite{Bauer:2001yt,Bauer:2003pi, Bosch:2004th}  
and including power corrections 
\cite{Lee:2004ja,Bosch:2004cb,Beneke:2004in}.
For inclusive $B$ decay these proofs are rather
straightforward.  Inclusive decay deals
with interactions between hard-collinear particles fluctuating
at the jet scale $m_b\Lambda_{\rm QCD}$ 
with soft particles fluctuating at the non-perturbative scale
$\Lambda_{\rm QCD}^2$.  The leading-order 
Lagrangian interactions between soft and 
hard-collinear particles can be 
decoupled by field redefinitions involving
Wilson lines \cite{Bauer:2001yt}.  
After integrating out hard fluctuations in a first 
step of matching, 
the factorization of the SCET matrix elements into a convolution
of jet and soft functions 
is  more or less a natural consequence of this
decoupling  at the level of the Lagrangian.

Applications of SCET to exclusive decay are
considerably more complicated 
\cite{Hill:2002vw, Bauer:2002aj,
Beneke:2003pa, Lange:2003pk}.
Exclusive processes typically involve both soft and
collinear particles fluctuating 
at the scale $\Lambda_{\rm QCD}^2$, in
addition to hard-collinear fluctuations
which are integrated out in the first step
of a two-step matching procedure. It has been argued that a  
low-energy theory of soft and collinear 
particles  contains a third mode, referred
to as soft-collinear \cite{Becher:2003qh}. This follows from
an analysis of loop diagrams with soft and collinear external
lines by the method of regions \cite{Beneke:1997zp}.
This soft-collinear ``messenger mode'' has the special property that
it can interact with both soft and collinear particles
without taking them far off shell. These modes introduce an 
additional, highly non-perturbative soft-collinear scale 
$\Lambda_{\rm QCD}^3/Q$.
To prove  factorization formulas of the type in 
(\ref{eq:fact}), one must show that the effects
of this fourth scale are irrelevant to the low-energy
matrix elements defining the soft functions $S$.  
This has been emphasized in  
\cite{Lange:2003pk, Becher:2003kh, Becher:2005fg}.
In $B$ decay the soft-collinear scale is
relevant at the endpoints of  convolution integrals
linking non-perturbative 
soft and collinear functions, so the soft-collinear field has 
often been associated with endpoint divergences
in these integrals \cite{Lange:2003pk}.

In this paper we show that the soft-collinear mode is 
relevant to an analysis of DIS at large $x$. 
Near the endpoint, DIS involves the three widely separated
scales $Q^2\gg (1-x)Q^2 \gg \Lambda_{\rm QCD}^2$.  Our main
finding is that we cannot correlate the two small scales by the
definition $\lambda^2 \sim (1-x)\sim \Lambda_{\rm QCD}/Q$ without
introducing a fourth scale, $ \Lambda_{\rm QCD}^3/Q
\sim Q^2 \lambda^6$.  
The appearance of this fourth scale is associated with
the soft-collinear mode. For values of $x$ satisfying
$1-x\sim \Lambda_{\rm QCD}/Q$,  the 
low-energy matrix element defining 
the parton distribution function 
involves fluctuations at both the collinear and soft-collinear
scales. An attempt to use effective field theory methods to prove
a factorization formula such as (\ref{eq:fact}) leads instead
to an expression 
\be\label{eq:nonfact}
H\left(\frac{Q^2}{\mu^2}\right) 
J\left(\frac{Q^2(1-x)}{\mu^2}, \frac{Q \omega_{sc}}{\mu^2} \right)\otimes
f\left( \frac{\Lambda_{\rm QCD}^2}{\mu^2}, 
\frac{\Lambda_{\rm QCD}^2 \omega_{sc}}{Q \mu^2}\right),
\ee 
where $\omega_{sc}\sim \Lambda_{\rm QCD}$ is a convolution variable.
Since the parton distribution function $f$ contains a non-perturbative
dependence on the large energy $Q$, factorization is spoiled. 

The organization of this paper is as follows.  In Section
\ref{sec:regions} we define our power counting and identify
the relevant SCET fields by applying the method of regions
to a representative loop diagram.  Section \ref{sec:SCET}
deals with matching the QCD Lagrangian and electromagnetic
current onto a version of SCET which accounts for 
these momentum regions.  In Section \ref{sec:tree} we show
with a tree-level example that the parton distribution function 
is sensitive to soft-collinear effects,
and discuss this further in  Section \ref{sec:oneloop} 
with a  one-loop calculation.  In
Section \ref{sec:sc} we summarize the implications of the 
soft-collinear mode on factorization.
We compare our results with previous work in
Section \ref{sec:comparison} and conclude in Section
\ref{sec:conclusions}.

\section{Power counting and momentum regions}
\label{sec:regions}

Deep inelastic scattering involves
the scattering of an energetic virtual photon with a large
invariant mass $q^2=-Q^2$ 
off a  proton with momentum $P$ to form 
a hadronic jet carrying momentum $p_x$ and 
an invariant mass  $p_x^2= Q^2 (1-x)/x$, where 
\be
x=-\frac{q^2}{2 P \cdot q}=\frac{Q^2}{2 P\cdot q}.
\ee
We are interested in the region of phase space  
where the hadronic jet carries a 
large energy of order $Q$, but has a small invariant mass on the
order of the jet scale $Q \Lambda_{\rm QCD}$. More precisely,
we work in the kinematic region  where
$p_x^2\sim Q \Lambda_{\rm QCD} \sim Q^2 (1-x)$.  
This  correlates 
the two small scales $1-x\sim \Lambda_{\rm QCD}/Q$. 
We make this explicit in the effective theory by introducing
an expansion parameter $\lambda^2\sim(1-x)\sim \Lambda_{\rm QCD}/Q$.
We then calculate the cross section as a double series in the
perturbative coupling constant and $\lambda$.  
In terms of $\lambda$ the 
invariants $P \cdot p_x\sim Q^2,\, p_x^2\sim Q^2 \lambda^2,$
and $P^2 \sim Q^2 \lambda^4$ define three widely separated
scales $Q^2 \gg Q^2\lambda^2 \gg Q^2\lambda^4$. 
In this paper we investigate  
whether we can derive a factorization formula which separates
the physics from these scales.
     
Our analysis relies on soft-collinear effective theory.
Unlike in applications of SCET to $B$ decay, there is no natural
Lorentz frame in which to describe the scattering process.
We find the Breit frame most convenient for what follows. 
In terms of two light-like vectors 
$n_{\pm}$ satisfying $\np\nm=2$, the  components
of the photon momentum $q^\mu$ in the Breit frame are given by
$(\np q, q_\perp,\nm q)=(-Q,0,Q)$.  If the 
proton momentum is  $P = (Q/x,0,M_p^2 x/Q)$,
then at leading order in $\lambda$ the jet momentum $p_x=q+P$ is given by 
$p_x=(Q(1-x)/x,0,Q)$ and satisfies $p_x^2=Q^2 (1-x)/x$.
We will refer to momenta with the scaling 
$p_c\sim Q(1,\lambda^2,\lambda^4)$
as collinear, and momenta with the scaling 
$p_{\hcbar}\sim Q(\lambda^2, \lambda, 1)$
as anti-hard-collinear.  With this terminology,
the proton momentum is collinear and the final-state
jet momentum is anti-hard-collinear.
Above and in the rest of the paper we work in
the reference frame where the transverse components
of the external momenta vanish. 

To construct the effective theory we must first identify
the momentum regions that produce on-shell
singularities in loop diagrams.  
SCET fields are then introduced to reproduce the 
effects of these momentum regions.  The relevant momentum 
regions depend on the choice of $x$.  In the kinematical 
regime where $1-x\sim\Lambda_{\rm QCD}/Q$, 
we find that we must consider
hard, anti-hard-collinear, collinear, and 
soft-collinear regions.  

The appearance of soft-collinear instead of soft modes will have
important consequences in our analysis.  Before we begin, it is
useful to explain their origin in simple terms. 
For the final-state jet to be anti-hard-collinear
requires that  $\np P+ \np q=\np p_x \sim Q \lambda^2$.  This is possible
only if $\np P =-\np q + \omega$, where $\omega$ is 
a residual momentum scaling as $\omega\sim Q\lambda^2$ and
$-\np q=Q$ is a large kinematic piece. 
This is similar to HQET, where
the $b$-quark momentum is $m_b v +k_s$, with $k_s$ a
soft residual momentum  $k_s\sim m_b(\lambda^2,\lambda^2,\lambda^2)$
and $m_b v$ a large kinematic piece.
For a collinear particle, however, the residual momentum
cannot be soft because $\nm k_s\sim Q\lambda^2$,
while for a collinear momentum  $\nm p_c \sim Q\lambda^4$. 
The simplest possibility is that the residual momentum 
scales as $Q(\lambda^2,\lambda^3,\lambda^4)$,
a scaling to which we will refer as soft-collinear. We 
will show below that this is indeed the scaling which is relevant
in loop diagrams.  This leads us to interpret the soft-collinear
mode as the residual momentum of a collinear field.   
  
\begin{figure}[t]
\begin{tabular}{c}
\hspace{4cm}\includegraphics[width=.4\textwidth]{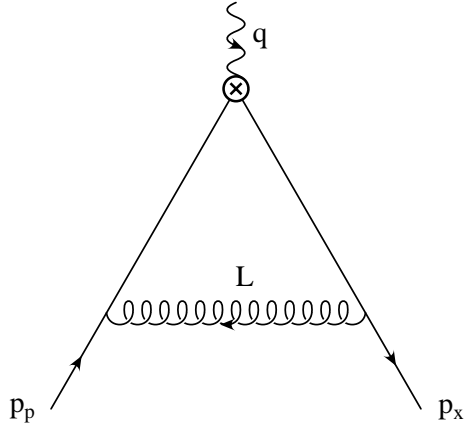} 
\end{tabular}
\caption{\label{fig:scalar}The triangle diagram. The momentum
$p_p$ is collinear and the momentum $p_x$ is anti-hard-collinear.}
\end{figure}

We will now make these  observations more rigorous
by analyzing a loop diagram using
the method of regions \cite{Beneke:1997zp}, 
similarly to \cite{Becher:2003qh}.
As a simplification, we begin with the scalar version of
the triangle diagram shown in
Figure \ref{fig:scalar}. This allows us
to identify the relevant momentum regions without complications related
to Dirac algebra. The external lines carry a collinear momentum
$p_p$ and an  anti-hard-collinear momentum $p_x=p_p+q$.
We set all masses to zero, and regularize  
IR divergences by keeping the external lines off shell
by an amount $p_x^2\sim  Q^2 \lambda^2$ and 
$p_p^2\sim P^2\sim  Q^2 \lambda^4$. 
The integral in the full theory is given by  
\bea\label{eq:fullintegral}
I& =& \int [dL]\frac{1}{(L+p_x)^2}\frac{1}{(L+p_p)^2}\frac{1}{L^2} \nl
&=&\frac{1}{Q^2}\left[\ln\frac{-p_x^2}{Q^2}\ln\frac{-p_p^2}{Q^2}
+\frac{\pi^2}{3}\right],
\eea
where we have defined the measure as 
\be
[dL]=i 16 \pi^2 \left(\frac{\mu^2 e^{\gamma_E}}{4\pi}\right)^\eps 
\frac{d^d L}{(2\pi)^d},
\ee
and expanded the  result to leading order
in $\lambda$. At leading order $Q^2= \np p_p \nm p_x$.

We seek to reproduce this result by the method of regions.
This strategy splits the loop integration into contributions
from momentum regions according to the scaling of their light-cone
components with $\lambda$.  The integrand is expanded as
appropriate for the particular momentum region before 
evaluating the integral. 
Once all relevant regions are identified, their sum reproduces
the full theory result. We start with the hard region, 
where the loop momentum scales as $L\sim Q(1,1,1).$ 
The expanded
integral is
\bea
I_h &=&\int [dL]\frac{1}{(L^2 + \nm p_x \np L)}
\frac{1}{(L^2 + \np p_p \nm L)}
\frac{1}{L^2}\nonumber \\
&=& \frac{1}{Q^2}\left[\frac{1}{\eps ^2}-
\frac{1}{\eps}\ln\frac{Q^2}{\mu^2}+
\frac{1}{2}\ln^2\frac{Q^2}{\mu^2}-\frac{\pi^2}{12}\right].
\eea 
We have regularized additional divergences with dimensional
regularization in $d=4-2 \eps$ dimensions.

The integral $I_h$ contains logarithms depending on the hard scale
$Q^2$. This is a generic feature: 
the result for a given region always involves logarithms 
at that momentum scale. For this reason 
we need to consider the anti-hard-collinear and collinear
regions, since these integrals can depend on $p_x^2$
and $p_p^2$.
For the anti-hard-collinear region, where  the 
loop momentum scales as $L\sim Q(\lambda^2,\lambda,1)$,
we find 
\bea
I_{\hcbar}&=&\int [dL]\frac{1}{(L+p_x)^2}
\frac{1}{(\nm L \np p_p)}\frac{1}{L^2} \nonumber \\ 
&=& \frac{1}{Q^2}\left[-\frac{1}{\eps ^2}+
\frac{1}{\eps}\ln\frac{-p_x^2}{\mu^2}-
\frac{1}{2}\ln^2\frac{-p_x^2}{\mu^2}+\frac{\pi^2}{12}\right].
\eea 
For the collinear region, where 
$L\sim Q (1,\lambda^2,\lambda^4)$, we have
\bea
I_{c}&=& \int[dL]\frac{1}{(\np L \nm p_x)}
\frac{1}{(L+p_p)^2}\frac{1}{L^2}\nonumber \\
&=& \frac{1}{Q^2}\left[-\frac{1}{\eps ^2}+
\frac{1}{\eps}\ln\frac{-p_p^2}{\mu^2}-
\frac{1}{2}\ln^2\frac{-p_p^2}{\mu^2}+\frac{\pi^2}{12}\right].
\eea 
Taking the sum of the regions considered so far does not
reproduce the result for the full integral  
(\ref{eq:fullintegral}). It is easy to check that
$I-I_h-I_{\hcbar}-I_c$ contains logarithms depending
on $p_x^2 p_p^2/Q^2\sim Q^2 \lambda^6$.  This is taken
into account by including the soft-collinear region, where 
$L\sim Q(\lambda^2,\lambda^3,\lambda^4)$.  This region
gives
\bea\label{eq:SCscalar}
I_{sc}&=& \int[dL]\frac{1}{(\np L \nm p_x + p_x^2)}
\frac{1}{(\nm L \np p_p +p_p^2 )}\frac{1}{L^2} \nonumber \\
&=& \frac{1}{Q^2}\left[\frac{1}{\eps ^2}+
\frac{1}{\eps}\ln\frac{Q^2\mu^2}{p_x^2 p_p^2}+
\frac{1}{2}\ln^2\left(\frac{p_x^2p_p^2}{Q^2\mu^2}\right)
+\frac{\pi^2}{4}\right].
\eea
Adding $I_h+I_{\bar{hc}}+I_c+I_{sc}$, 
we see that the poles cancel, and that we recover the 
result for the full integral given in (\ref{eq:fullintegral}).
We will  construct a version of SCET which 
accounts for these momentum regions in the next section.

Note that the soft and hard-collinear regions  are 
needed in applications of SCET to $B$ decay, but are
not needed here.     
The soft region, where $L\sim Q(\lambda^2,\lambda^2,\lambda^2)$,
is irrelevant because
\bea\label{eq:Is}
I_s &=& \int[dL]\frac{1}{(\np L \nm p_x + p_x^2)}
\frac{1}{(\nm L \np p_p)}\frac{1}{L^2} \nl
 &=& \frac{1}{Q^2}\int[dL]\frac{1}{(\np L  + \np p_x)}
\frac{1}{(\nm L)}\frac{1}{L^2}=0.
\eea
To derive the second line we used 
$p_x^2=\np p_x \nm p_x$ (recall $p_{x_\perp}=0$), and 
then that scaleless integrals vanish in dimensional 
regularization. The hard-collinear integrand,
where $L\sim Q(1,\lambda, \lambda^2)$, is also scaleless and 
vanishes. 

While it may be possible to eliminate the soft-collinear
scale $\Lambda_{\rm QCD}^3/Q$ by introducing an
IR regulator to cut off momentum regions with 
virtuality smaller than the QCD scale $\Lambda_{\rm QCD}^2$,
we find it more convenient to keep the collinear 
quarks off shell by an amount $p_p^2\sim P^2$ 
and use dimensional regularization.  
In our end analysis we will adopt the philosophy
of  \cite{Becher:2003kh, Becher:2005fg}, and 
interpret any sensitivity of low-energy matrix 
elements to the soft-collinear 
mode as a breakdown of factorization.

We should emphasize that all results are frame
independent. It is also possible to carry out 
the analysis in the target rest frame, where the proton
momentum is soft.  We can identify the scaling of the light-cone
components of the momentum regions in the rest frame 
by performing a Lorentz boost to this frame,
which amounts to rescaling $n_{\pm}$. 
The components of a generic momentum change
according to $(\np p, p_\perp ,\nm p)\to (\np p \lambda^2,
p_\perp, \nm p \lambda^{-2})$.  The correspondence between
the two frames is given by

\begin{center}
\begin{tabular}{lllll}
 &&&\\ & &   Breit Frame &   & Rest Frame  \\
 &hard & $Q(1,1,1)$ & \hspace{.5cm}  
$\leftrightarrow$ \hspace{.5cm} 
& $Q(\lambda^2,1,\frac{1}{\lambda^2})$ \\
&anti-hard-collinear & $Q(\lambda^2,\lambda,1)$&  
\hspace{.5cm}  
$\leftrightarrow$ \hspace{.5cm} & 
$Q(\lambda^4,\lambda,\frac{1}{\lambda^2})$\\
&collinear & $Q(1,\lambda^2,\lambda^4)$ &  \hspace{.5cm}  
$\leftrightarrow$ \hspace{.5cm} & 
$Q(\lambda^2,\lambda^2,\lambda^2)$ \\
& soft-collinear  & $Q(\lambda^2,\lambda^3,\lambda^4)$ 
& \hspace{.5cm}  
$\leftrightarrow$ \hspace{.5cm} & 
$Q(\lambda^4,\lambda^3,\lambda^2)$\\
 & & &
\end{tabular}
\end{center}
Although the individual light-cone components of the momentum 
regions scale differently in the two frames,
the number of regions is the same.  
Moreover, the result for each region depends on invariants at that
scale and is therefore frame independent.  
This can be seen from the explicit results, or by noticing
that each integrand is invariant under the simultaneous 
rescalings of $n_{\pm}$ shown above.  In the effective theory this 
is referred to as reparameterization invariance (RPI) 
\cite{Manohar:2002fd, Chay:2002vy}.

\section{Matching onto SCET}\label{sec:SCET}
This section deals with matching the QCD Lagrangian and 
electromagnetic current onto SCET.  Our eventual goal is to examine the 
factorization properties of 
the hadronic tensor using effective field theory methods.
In inclusive processes all QCD effects are contained
in the hadronic tensor, which is given by the spin-averaged
matrix element between proton states
\be\label{eq:htensor}
W^{\mu\nu}=\frac{1}{\pi}{\rm Im}\langle P|T^{\mu\nu}|P\rangle,
\ee
where the current correlator $T^{\mu\nu}$ is defined through 
the time-ordered product
\be\label{eq:correlator}
T^{\mu\nu}= i\int d^4 z e^{iq z}{\rm T}
\left\{J^{\mu\dagger}(z)J^\nu(0)\right\}.
\ee
Here $J^\mu$ is the electromagnetic current, and $q$ is the 
momentum of the incoming photon.  
We will evaluate the correlator in effective field
theory by separating the contributions from the momentum
regions identified in the previous section, namely
\begin{center}
\begin{tabular}{ll}
hard & \hspace{.5cm} $Q(1,1,1)$ \\
anti-hard-collinear & \hspace{.5cm} $Q(\lambda^2,\lambda,1)$ \\
collinear &\hspace{.5cm} $Q(1,\lambda^2,\lambda^4)$ \\
soft-collinear & \hspace{.5cm} $Q(\lambda^2,\lambda^3,\lambda^4)$
\end{tabular}
\end{center}
We calculate the hadronic tensor using a two-step matching
procedure familiar from applications of SCET to inclusive $B$ decay
in the shape-function region \cite{Bauer:2001yt, Bauer:2003pi, Bosch:2004th}. 
In the first step, we  match the QCD Lagrangian and electromagnetic
current onto SCET by integrating out fluctuations 
at the hard scale $Q^2$ and introducing effective theory
fields for the regions $p_{\bar {hc}},\,p_c,p_{sc}$.
The Lagrangian can be derived exactly, and
will be discussed in the next sub-section.  The current, on the
other hand, receives corrections from fluctuations at the hard
scale.  These corrections can be absorbed into a hard Wilson
coefficient, which we will calculate at one loop
in Section \ref{sec:currents}.
In a second step of matching we evaluate the hadronic tensor
(\ref{eq:htensor}) using the SCET Lagrangian
and current.  In this step
of matching we integrate out fluctuations at the hard-collinear
scale $Q\Lambda_{\rm QCD}$ and match onto the parton 
distribution function.  We discuss
this at tree level in Section \ref{sec:tree} and at 
one loop in Section \ref{sec:oneloop}.

\subsection{SCET Lagrangian}\label{sec:lagrangians}
The QCD Lagrangian for light quarks contains no hard scale and the 
SCET Lagrangian can be derived exactly \cite{Beneke:2002ph}.  
For the case at hand, we have 
\be
{\cal L}_{\rm QCD}\to {\cal L}_{c+sc} + {\cal L}_{\hcbar +sc} +{\cal L}_{sc} +{\cal L}_{YM},
\ee
where ${\cal L}_{c+sc}$ contains the collinear Lagrangian as well
as interactions with the soft-collinear gluon field, and
analogously for ${\cal L}_{\hcbar + sc}$. There
is no interaction term ${\cal L}_{c+\hcbar}$ for processes
where the initial and final states contain only one type of
collinear field \cite{Becher:2005fg}. The soft-collinear
Lagrangian ${\cal L}_{sc}$ can be found in \cite{Becher:2003qh},
and the Yang-Mills Lagrangian for each sector is the same
as in QCD.

The Lagrangian ${\cal L}_{c+sc}$
can be derived using the methods of
\cite{Beneke:2002ph, Beneke:2002ni}, as was done in 
\cite{Becher:2003qh}.   
The result for the leading-order Lagrangian ${\cal L}_{c+sc}$
is
\bea\label{eq:Lssc}
{\cal L}_{c+sc}= \bar \xi_c\left(i\nm D_{c+sc}+(i\Slash{D}_{c_\perp }-m_q)
\frac{1}{i\np D_c}(i\Slash{D}_{c_\perp }+m_q) \right)
\frac{\slash{n}_+}{2}\xi_c,
\eea
where $i D_{c+sc}^\mu=i\partial^\mu + g A_c^\mu +g A_{sc}^\mu$.
In interactions between collinear and 
soft-collinear fields the soft-collinear fields
are multipole expanded and depend  on $z_-^\mu=(\np z)\nm^\mu/2$.
We have omitted a pure glue interaction term, which will not be
needed here.
We can derive the Lagrangian  ${\cal L}_{\hcbar+sc}$
by making the replacements $\nm \leftrightarrow \np$
and $\phi_c\to \phi_{\hcbar}$ in the expressions above. 
The result is 
\bea\label{eq:Lhcbarsc}
{\cal L}_{\hcbar+sc}= \bar \xi_{\hcbar}
\left(i\np D_{\hcbar+sc}+i\Slash{D}_{\hcbar_\perp}
\frac{1}{i\nm D_{\hcbar}}i\Slash{D}_{ \hcbar_\perp } \right)
\frac{\slash{n}_-}{2}\xi_{\hcbar}.
\eea 
We have again omitted a pure glue interaction term. 
In interactions between anti-hard-collinear and soft-collinear fields
the soft-collinear fields must be multipole expanded and depend only on
$z_+^\mu=(\nm z)\np^\mu/2$.

We have included a collinear quark mass  
$m_q\sim \Lambda_{\rm QCD}\sim Q\lambda^2$ in the leading-order
Lagrangian ${\cal L}_{c+sc}$ above. 
We are free to include such a mass without changing
the regions analysis. In fact, keeping the collinear momentum 
off shell by an amount $p_p^2\sim Q^2\lambda^4$ 
effectively gave such a scale to the collinear line, adding an 
actual mass just changes $p_p^2\to p_p^2-m_q^2$ in the collinear
propagator.  This does not eliminate soft-collinear effects.  We
checked this claim by modifying the scalar triangle integral
to include a mass $m_q\sim Q\lambda^2$ 
for the collinear line and confirmed
that, at least to one loop, the regions analysis 
is unchanged.  We have no proof that the regions 
analysis is unchanged beyond one loop, and in the following
calculations we will always set $m_q=0$ for simplicity. 

A property of the Lagrangians crucial for 
factorization proofs is that the 
soft-collinear fields can be decoupled from 
the anti-hard-collinear
and collinear fields through field redefinitions involving Wilson 
lines \cite{Becher:2003qh}. We introduce the Wilson lines
\bea
S_{sc}(z)&=&{\rm P\, exp}\left(ig\int_{-\infty}^0 
ds\,\nm A_{sc}(z+s\nm)\right)\\
S_{\scbar}(z)&=&{\rm P\, exp}\left(ig\int_{-\infty}^0 
ds\,\np A_{sc}(z+s\np)\right)
\eea
along with similar objects $W_c$ and $W_{\hcbar}$, 
where the soft-collinear fields are replaced by 
collinear or anti-hard-collinear fields, and 
$\np \leftrightarrow \nm$. 
After making the field redefinitions
\bea\label{eq:decoupling}
\xi_c&=& S_{sc}\xi^{(0)}_c,\quad A_c= S_{sc}A_c^{(0)}S_{sc}^\dagger,
\quad W_c= S_{sc}W_c^{(0)} S_{sc}^\dagger, \\
\xi_{\hcbar}&=& S_{\scbar}\xi_{\hcbar}^{(0)},\quad 
A_{\hcbar}= S_{\scbar}A_{\hcbar}^{(0)}S_{\scbar}^\dagger,
\quad W_{\hcbar}= S_{\scbar}W_{\hcbar}^{(0)} S_{\scbar}^\dagger,
\nonumber
\eea
the fields with the superscript $0$ no longer interact with
the soft-collinear fields.  This factorization of soft-collinear
fields at the level of the Lagrangians does not 
guarantee the factorization of the current correlator 
(\ref{eq:correlator}), however,
because the effects may reappear in time-ordered products
with the external currents \cite{Becher:2003kh}.

\subsection{SCET  current at one loop}
\label{sec:currents}
Having obtained the relevant SCET Lagrangian, we now 
consider the one-loop matching of 
the electromagnetic current onto its effective
field theory expression.  This was done previously in
\cite{manohar} and we agree with the results obtained there.
We will repeat the calculation to show how logarithms 
related to the soft-collinear mode are essential to the
analysis. 

At leading order the matching of the electromagnetic current
onto SCET takes the form
\be\label{eq:falsch}
\bar \psi_c(z)\gamma^\mu \psi_{\hcbar}(z)\to 
\int  d s  d t \, \tilde C( s, t, \mu)
(\bar\xi_c W_c)(z+s\np)
\gamma^\mu (W_{\hcbar}^\dagger \xi_{\hcbar})(z+t\nm).
\ee
As in \cite{manohar}, we consider a single quark flavor
with unit charge.  The convolution arises because $\np p_c$ 
and $\nm p_{\hcbar}$ are on the
order of the hard scale, so the operator can 
be non-local by an amount $1/Q$ in these directions.  
Setting $z$ to zero and using translational invariance, the
current can be written as 
\be
C(\np P_c \,\nm P_{\hcbar},\mu)(\bar\xi_c W_c)(0)
\gamma^\mu (W_{\hcbar}^\dagger \xi_{\hcbar})(0),
\ee
where the Fourier-transformed coefficient function is
\be
C(\np P_c\,\nm P_{\hcbar} ,\mu)=\int ds dt \, 
\tilde C( s, t)e^{i(s\np P_c- t\nm P_{\hcbar})},
\ee
and  $P_{c,\hcbar}$ are momentum operators. In our case
these are both $Q$  so we have $C(Q^2,\mu)$.   

\begin{figure}[t]
\begin{tabular}{c}
\hspace{0cm}\includegraphics[width=1\textwidth]{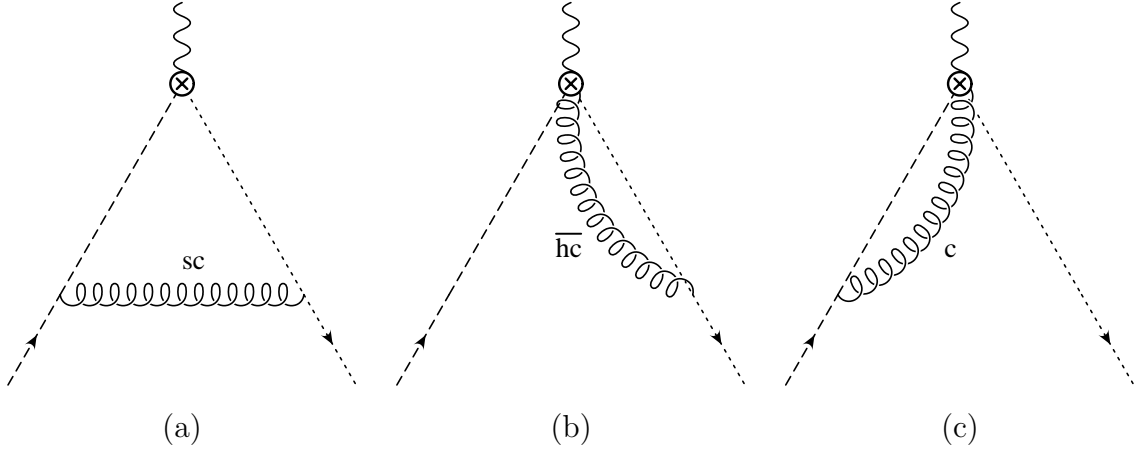} 
\end{tabular}
\caption{\label{fig:SCETcurrent}One-loop corrections to the SCET current.  
The long-dashed lines are collinear and the short-dashed lines
anti-hard-collinear. The gluon scaling is indicated explicitly.}
\end{figure}

To calculate the one-loop matching conditions we take 
the difference of the  QCD result from
that evaluated in SCET. The QCD graph is the same as in
Figure \ref{fig:scalar} but evaluated with the Feynman 
rules of QCD. We find it useful to break up the QCD result into 
contributions from each momentum region, as we 
did with the scalar triangle. The matching conditions
are related only to the hard region. For the QCD result
we find
\bea
I_{\rm QCD}&=&\frac{C_F\alpha_s}{4\pi}\gamma^\mu\left[\frac{1}{\eps_{\rm UV}}
-\ln\frac{Q^2}{\mu^2}-2\ln\frac{-p_p^2}{Q^2}\ln\frac{-p_x^2}{Q^2}-
2\ln\frac{-p_p^2}{Q^2}-2\ln\frac{-p_x^2}{Q^2}-\frac{2\pi^2}{3}\right]\nl
&=&I_h + I_{\hcbar}+ I_c + I_{sc},
\eea
where   
\bea
I_h&=& \frac{C_F \alpha_s }{4\pi}\gamma^\mu
\left[-\frac{2}{\eps^2} +
\frac{2}{\eps}\left(\ln\frac{Q^2}{\mu^2}-2\right)
+\frac{1}{\eps_{\rm UV}} - \ln^2 \frac{Q^2}{\mu^2} +
3\ln\frac{Q^2}{\mu^2}
+\frac{\pi^2}{6} -8 \right],
\label{eq:QCDhard}\\
I_{\hcbar}&=& \frac{C_F\alpha_s}{4\pi} \gamma^\mu
\left[\frac{2}{\eps^2}-\frac{2}{\eps}\left(
\ln\frac{-p_x^2}{\mu^2}-1\right)+\ln^2\frac{-p_x^2}{\mu^2}- 
2\ln\frac{-p_x^2}{\mu^2}-\frac{\pi^2}{6}+4\right],
\label{eq:QCDhcbar} \\
I_c&=& \frac{C_F\alpha_s}{4\pi} \gamma^\mu
\left[\frac{2}{\eps^2}-\frac{2}{\eps}\left(
\ln\frac{-p_p^2}{\mu^2}-1\right)+\ln^2\frac{-p_p^2}{\mu^2}- 
2\ln\frac{-p_p^2}{\mu^2}-\frac{\pi^2}{6}+4\right],
\label{eq:QCDc}\\
I_{sc}&=& \frac{C_F\alpha_s}{4\pi} \gamma^\mu
\left[-\frac{2}{\eps^2}+\frac{2}{\eps}\ln\frac{p_x^2 p_p^2}{Q^2\mu^2}
-\ln^2\frac{p_x^2 p_p^2}{Q^2\mu^2}-\frac{\pi^2}{2}\right]
\label{eq:QCDsc}.
\eea
We have expanded all  results to leading order 
in $\lambda$, and used that $Q^2=\np p_p \nm p_x$ at 
this order. 
We must supplement these graphs with the wave-function
renormalization for off-shell quarks, which gives 
a contribution 
\be\label{eq:wave}
I_w=\frac12\frac{C_F\alpha_s}{4\pi} \gamma^\mu
\left[-\frac{1}{\eps_{\rm UV}}-1+ \ln \frac{-p_i^2}{\mu^2}\right]
\ee 
for each external quark line.
The UV poles cancel in the sum $I_h + I_w$, as required by
current conservation.

The next step is to evaluate the
SCET diagrams in Figure \ref{fig:SCETcurrent}.
Evaluating the graphs in the figure using
the Feynman rules of SCET  reproduces the result for the 
QCD regions calculation.  By this we mean that Figure 2(a)
evaluates to $I_{sc}$, Figure 2(b) to $I_{\hcbar}$,
and Figure 2(c) to $I_c$.  This just confirms that
we have constructed the effective theory correctly. 
The wave-function graphs in the effective theory are the
same as  (\ref{eq:wave}) \cite{Bauer:2000ew}.

The difference
between the two theories is that the hard integral $I_h$ is absent in
SCET. Its finite part is taken into account by
a hard matching coefficient, and its infinite part is reproduced
by a renormalization factor $Z_J$ applied to the bare current. 
Including the tree-level contribution, the 
matching coefficient is therefore 
\be
C(Q^2,\mu)=1+\frac{C_F\alpha_s}{4\pi}\left[-\ln^2\frac{Q^2}{\mu^2}
+3\ln\frac{Q^2}{\mu^2}+\frac{\pi^2}{6}-8\right],
\ee
and the renormalization factor is 
\be\label{eq:ZJ}
Z_J=1+\frac{C_F\alpha_s}{4\pi}\left[-\frac{2}{\eps^2}-\frac{3}{\eps}+
\frac{2}{\eps}\ln\frac{Q^2}{\mu^2}\right].
\ee
The hard coefficient $C(Q^2,\mu)$ and the renormalization factor
$Z_J$ depend on the hard scale  $Q^2$. 
For the infinite counter terms,
this is possible only after a cancellation between logarithms that
occurs when adding the anti-hard-collinear, collinear, and
soft-collinear graphs. That logarithms of UV origin related to the
soft-collinear field are needed to ensure that the renormalization
factor $Z_J$ depends only on the hard scale $Q^2$ was first 
noted in \cite{Becher:2003kh}, in a slightly different context.
On the other hand, no such cancellation occurs for the finite 
terms, where the sum of the anti-hard-collinear, collinear, and soft
collinear graphs still contains logarithms at each scale.
We will see in Sections \ref{sec:oneloop} and \ref{sec:sc} 
that the matrix element 
of the current correlator shares this property, and that the
logarithms related to the soft-collinear scale 
cause problems for factorization.

\section{Matching onto parton distributions at tree level}
\label{sec:tree}
Matching onto the intermediate theory has absorbed the effects
of hard fluctuations into a short-distance Wilson coefficient. This leaves
the three widely separated scales 
$Q \Lambda_{\rm QCD}\gg  \Lambda_{\rm QCD}^2 \gg \Lambda_{\rm QCD}^3/Q$.
We now examine the factorization properties of the hadronic tensor 
(\ref{eq:htensor}). To achieve a perturbative 
factorization of the form (\ref{eq:fact}), we would
need to show that the soft-collinear scale $\Lambda_{\rm QCD}^3/Q$
is irrelevant.  We could then perform a second and final step
of matching at the scale $Q \Lambda_{\rm QCD}$, and 
identify the associated matching 
coefficient with the jet function $J$. The low-energy matrix
element would define a parton distribution function $f$ 
characterized by fluctuations at the collinear scale 
$\Lambda_{\rm QCD}^2$ only, which would be linked to $J$ 
in a convolution integral. The purpose of this section is 
to demonstrate with a tree-level example that this 
is impossible, by showing that soft-collinear effects
do not decouple from the low-energy matrix element.  
In fact, the jet function is linked to the parton distribution
function by a convolution variable related to  
the soft-collinear scale.  We will argue that 
a full separation of scales would require integrating
out the collinear scale $\Lambda_{\rm QCD}^2$ in 
a third step of matching, which however cannot be 
done perturbatively.

\begin{figure}[t]
\begin{tabular}{c}
\hspace{0cm}\includegraphics[width=1\textwidth]{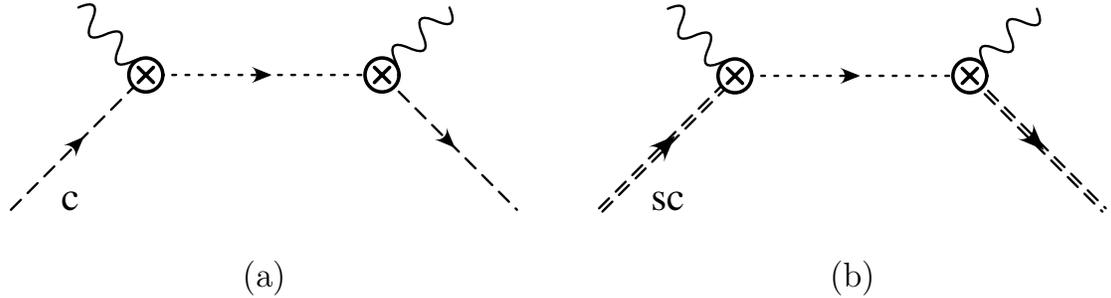} 
\end{tabular}
\caption{
Tree-level contribution to the current correlator in SCET,
evaluated in terms of collinear fields $\xi_c$ (a),
and soft-collinear fields $\xi_{Q,sc}$ (b). 
The short-dashed propagator is 
anti-hard-collinear.\label{fig:tree}}
\end{figure}

At leading order, the current correlator (\ref{eq:correlator})
is given by the time-ordered product 
\be
T^{\mu\nu}=i\int d^4 z e^{iq\cdot z}
{\rm T}\bigg\{\bar\chi_c(z)\gamma^\mu
 \chi_{\hcbar}(z)\bar\chi_{\hcbar}(0)
\gamma^\nu \chi_c(0)\bigg\},
\ee  
where we have defined the fields
\be
\chi_c\equiv W_c^\dagger\xi_c,\qquad 
 \chi_{\hcbar}\equiv W_{\hcbar}^\dagger\xi_{\hcbar},
\ee
which are manifestly gauge invariant under anti-hard-collinear 
and collinear gauge transformations.  At tree level
and to lowest order in $g$ the hard Wilson coefficient $C(Q^2,\mu)$ 
and the Wilson lines $W$ are unity.
We perform a second step of matching by integrating out 
the anti-hard-collinear fields. 
This is done at the scale $Q \Lambda_{\rm QCD}$, which we treat 
as perturbative.  To do this at tree level, we first
perform the decoupling redefinition (\ref{eq:decoupling})
on the anti-hard-collinear fields (and 
immediately drop the superscript (0)), and then contract
the anti-hard-collinear fields into a propagator. This is represented
by the Feynman diagram in Figure \ref{fig:tree}(a).
The anti-hard-collinear propagator is
given in momentum space by
\be
\langle 0|\xi_{\hcbar}(z)_{a\alpha}
\bar\xi_{\hcbar}(0)_{b\beta}|0\rangle=
\int \frac{d^4 L}{(2\pi)^4}e^{-iL z}\frac{i\nm L}{L^2 + i 0}
\left(\frac{\slash{n}_+}{2}\right)_{\alpha\beta}\delta_{ab}.
\ee
This forces $z$ to scale as an anti-hard-collinear quantity, and 
we need to perform the multipole expansion accordingly.  
This is in general different from the multipole expansion
in SCET Lagrangian interactions, because the photon injects a
large external momentum $q$ into the diagram. In particular, since
$z$ scales as anti-hard-collinear, the collinear and soft-collinear
fields can depend only on $z_+^\mu=(\nm z/2) \np^\mu$.   
The result for the current correlator is then
\bea\label{eq:int}
T^{\mu\nu}&=&-\int d^4 z e^{iq z}\bar\xi_c(z_+)\gamma^\mu
S_{\bar{sc}}(z_+)\frac{\slash{n}_+}{2}S^\dagger_{\bar{sc}}(0)
\gamma^\nu\xi_c(0) \nl && \int \frac{d^4 L}{(2\pi)^4}e^{-iL z}
\frac{\nm L}{\nm L \np L+L_\perp^2+i0}
\eea
The soft-collinear and collinear fields do not depend on $\np z$ or
$z_\perp$, so we can perform these integrations.  
We then have
\bea \label{eq:cctree}
T^{\mu\nu}&=&-\int \frac{d(\nm z)}{2} \frac{d (\np L)}{2\pi}
e^{-i\np L\nm z/2}\frac{\nm q}{\nm q \np L+i0}\nl &&
e^{i\np q\nm z/2}\bar\xi_c(z_+)S_{\scbar}(z_+)S^\dagger_{\scbar}(0)
\gamma^\mu\frac{\slash{n}_+}{2}\gamma^\nu\xi_c(0).
\eea
Even though $\np p_c \sim Q$ and $\np p_{sc}\sim Q\lambda^2$ 
we cannot set the argument of the soft-collinear Wilson
line $S_{\bar{sc}}(z_+)$ to zero. This is because 
$\np p_c + \np q\sim Q\lambda^2$, so we need to 
keep $\np p_{sc}\sim Q\lambda^2$ in the $\np L\sim Q\lambda^2$ 
component of the anti-hard-collinear propagator.  We will discuss
this further below.  For now, we simply note that soft-collinear 
effects do not decouple even at leading order 
in the $1/Q$ expansion.

In order to calculate the hadronic tensor (\ref{eq:htensor})
we now take the matrix element of the current 
correlator between proton states. We define a parton 
distribution function through the spin-averaged matrix element
\bea\label{eq:f}
&&\langle P|\bar\chi_c(t\np)S_{\scbar}(t\np)S^\dagger_{\scbar}(0)
\gamma^\mu\frac{\slash{n}_+}{2}\gamma^\nu\chi_c(0)|P\rangle 
=\tilde f(t){\rm tr}\left[\frac{\slash{n}_-}{2}
\gamma^\mu\frac{\slash{n}_+}{2}\gamma_\nu\right](-\np q).
\eea
The factor of $-\np q = Q + {\cal O}(Q\lambda^2)$ 
preserves manifest boost invariance. 
Although not necessary for tree-level matching,
we have reinserted the Wilson lines $W_c$
in order to define a gauge invariant hadronic matrix element.
The Fourier transformed function is 
\be
\tilde f(t)=
\int d\omega e^{-i\omega t} f(\omega).
\ee
Inserting this into (\ref{eq:int}), the  hadronic tensor becomes
\be\label{eq:factorized}
W^{\mu\nu}=-\frac{1}{\pi}{\rm Im}\int d\omega f(\omega)
\frac{Q}{\np q -\omega +i 0}
{\rm tr}\left[\frac{\slash{n}_-}{2}\gamma^\mu
\frac{\slash{n}_+}{2}\gamma^\nu\right].
\ee

As written,  (\ref{eq:factorized}) obscures the power
counting in the effective theory.
We have used a delta function to eliminate 
$\np L \sim Q \lambda^2$, so we must have  
$\np q -\omega \sim Q \lambda^2$.  
This requires that $\omega=\np q+ \omega_{sc}$, 
where $\omega_{sc}\sim Q\lambda^2$.  The large component
of the collinear momentum simply balances that of the incoming
photon momentum, it is the residual soft-collinear component
$\omega_{sc}$ which controls the dynamics.     
We can make this transparent by integrating out
the collinear scale $\Lambda_{\rm QCD}^2$ 
and matching onto a low-energy theory defined at the 
soft-collinear scale. 
This is of course not possible because QCD is already strongly
coupled at the collinear scale, and we will revisit
this point below.  At tree level, however, 
such a matching is trivial: we simply introduce a new field  
by writing $e^{i Q \nm z/2}\xi_c(z)
=\xi_{Q,sc}(z)$, so that 
$\xi_{Q,sc}$  carries the residual 
momentum $\omega_{sc}$.\footnote {This field is similar 
to the SCET field $\xi_{Q,n}$ 
defined in the label formalism \cite{Bauer:2001yt},
the difference being that the residual momentum 
is soft-collinear instead of ultra-soft.}  
Figure \ref{fig:tree}(b) shows the 
tree-level diagram for the
current correlator evaluated using the $\xi_{Q,sc}$.  
We calculate the correlator using the same steps as before,
and define a distribution function $f_{sc}$ 
analogously to (\ref{eq:f}), 
but in terms of $\xi_{Q,sc}$ instead of $\chi_c$. 
We furthermore choose $\np q= -Q + \np p_x$, 
with $\np p_x=Q(1-x)$.  The result is
\be\label{eq:tree}
W^{\mu\nu}=\int d\omega_{sc} \,J(\np p_x -\omega_{sc}) f_{sc}(\omega_{sc})
{\rm tr}\left[\frac{\slash{n}_-}{2}\gamma^\mu
\frac{\slash{n}_+}{2}\gamma^\nu\right],
\ee
where 
\be
J(\np p_x-\omega_{sc}) = -\frac{1}{\pi}{\rm Im}
\frac{Q}{\np p_x-\omega_{sc} + i0}=Q\delta(\np p_x -\omega_{sc}).
\ee
This completes the tree-level matching calculation.
We could have obtained this same result in the
free-quark decay picture by calculating
the diagram in Figure \ref{fig:tree} and taking the imaginary part.
In the free-quark picture at tree level we can interpret
$f_{sc}(\omega_{sc})= \delta(\omega_{sc})$, so that
the  convolution $J\otimes f_{sc}$ reproduces 
the result for the diagram.  We went through the 
extra step of defining the parton distribution function 
in terms of a hadronic matrix 
element in order to draw some parallels between
(\ref{eq:tree}) and the factorization formula 
derived for inclusive $B$ decay in the shape-function region,
where one finds a convolution of the form 
\cite{Korchemsky:1994jb,Bauer:2001yt}
\be
\int d\omega_s \,J(\np p_x-\omega_s)S(\omega_s),
\ee
with $\np p_x=m_b(1-x)\sim \Lambda_{\rm QCD}$. 
The function $J(\omega_s)$ is a perturbatively calculable jet function
containing  physics at the scale $m_b\Lambda_{\rm QCD}$,
and $S(\omega_s)$ is a shape function containing non-perturbative
effects at the scale $\Lambda_{\rm QCD}^2$. It is defined by
the HQET matrix element
\be
S(\omega_s)=\int dt e^{-it\omega_s}
\langle \bar B_v|\bar h_v(tn_-)\,h_v(0)|\bar B_v\rangle.
\ee 
The crucial difference between $B$ decay and DIS 
is that the heavy-quark field $h_v$ carries a residual 
momentum $\omega_s$ which 
is soft. Matching 
can be stopped at the perturbative scale 
$m_b\Lambda_{\rm QCD}$.  This should be compared
with (\ref{eq:tree}), where the convolution involves
the soft-collinear residual momentum $\omega_{sc}$.
To isolate the physics at this low scale requires
an extra step of matching at the scale
$p_p^2\sim \Lambda_{\rm QCD}^2$. We did this above in
order to define $f_{sc}$, but it is important to
understand that this was only a formal manipulation.  
Since we cannot do this
matching perturbatively, we  must always 
lump the collinear and soft-collinear effects together
into one non-perturbative function. 
We will emphasize in Section \ref{sec:sc} that 
any sensitivity of the parton distribution function
to the soft-collinear scale signals 
a breakdown of factorization.  
However, to explain how effective field theory could in 
principle be used to separate all the scales, 
we end this section by considering a fictitious QCD
where perturbation theory is valid at the collinear scale. 
In this fictitious theory we can remove 
collinear fluctuations when matching 
the electromagnetic current onto SCET.
This matching takes the form
\be \label{eq:richtig}
\bar\psi_c\gamma^\mu \psi_{\hcbar}\to 
 C(Q^2,\mu) D_c(p_p^2,\mu)
\bar\xi_{Q,sc}
\gamma^\mu  \chi_{\hcbar}.
\ee
The matching coefficient $D_c(p_p^2,\mu)$ 
reproduces the effects of collinear loop diagrams,
and could be obtained at one loop from the finite
part of our expressions in Section 
\ref{sec:currents}, see (\ref{eq:QCDhard}-\ref{eq:QCDsc}).  
To consider such a (fictitious) matching of the SCET 
current will be useful in some of the discussion in the
next two sections.

\section{Matching onto parton distributions at one loop}
\label{sec:oneloop}
\begin{figure}[t]
\begin{tabular}{c}
\hspace{0cm}\includegraphics[width=1\textwidth]{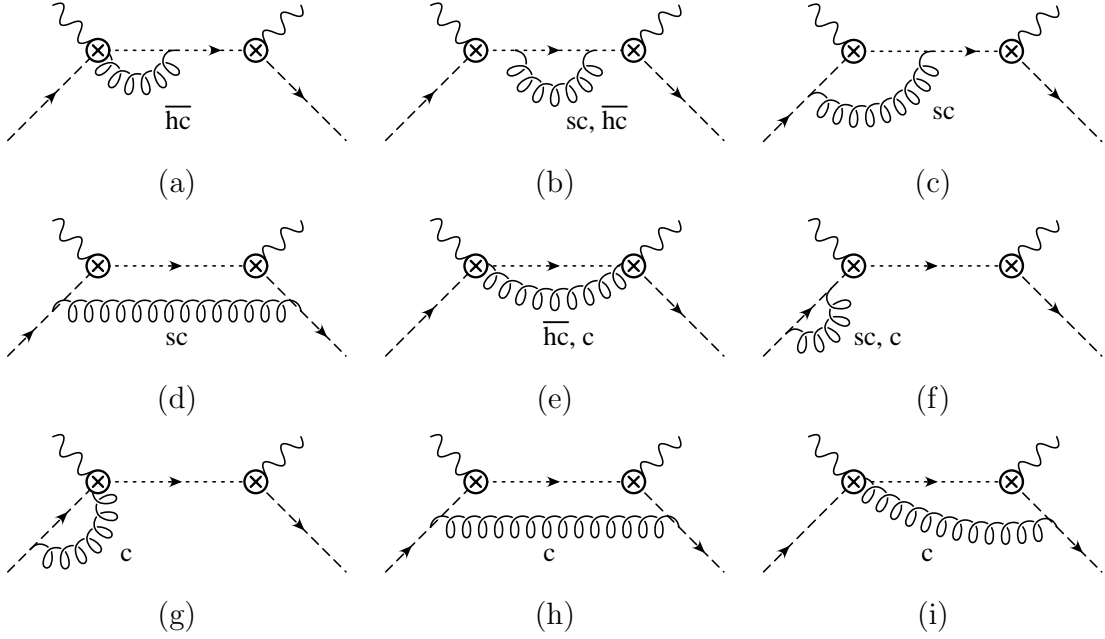} 
\end{tabular}
\caption{\label{fig:1loop}
One-loop corrections to the current correlator.  The long-dashed 
lines are collinear and the short-dashed lines 
anti-hard-collinear.  The gluon scaling is labeled explicitly.
The mirror image graphs are not shown. The part of (e)
involving collinear exchange, (h),  and (i) 
should {\it not} be included in the effective theory.}
\end{figure}

In this section we examine the one-loop corrections to the
current correlator (\ref{eq:correlator}),  and interpret
the results in terms of the effective theory.
The relevant one-loop diagrams  
are shown in  Figure \ref{fig:1loop}.  
Note that the graph in Figure \ref{fig:1loop}(e) 
containing collinear exchange, as well as graphs 
\ref{fig:1loop}(h) and \ref{fig:1loop}(i),  
are not actually SCET graphs.  
In these graphs the short-dashed propagator
is hard, not anti-hard-collinear. It was our 
intention to remove all hard fluctuations in the first
step of matching, but we have clearly not done so. Although
these graphs are power suppressed by a factor of $p_x^2/Q^2 \sim
\lambda^2$, we find it awkward 
to generate power-suppressed graphs from the leading-order
Feynman rules of the effective theory. 
A formal solution to this problem is to remove the collinear scale 
when matching the current, as in (\ref{eq:richtig}).
We cannot do this matching perturbatively, but  
we are interested only in the sum of collinear and soft-collinear
graphs, which can equally well be written in this way.  This 
simplifies the book-keeping, because after taking this step
the graphs \ref{fig:1loop}(e), \ref{fig:1loop}(h) 
and \ref{fig:1loop}(i) no longer exist, so the
short-dashed propagator is always anti-hard-collinear.    
Note that we have not drawn box diagrams 
related to gluon distributions. These are power suppressed, either 
because the intermediate propagator is hard, analogously to
\ref{fig:1loop}(h), or because 
they involve insertions of soft-collinear quark fields (not to 
be confused with $\xi_{Q,sc}$), which 
are absent from the leading order 
SCET Lagrangian \cite{Becher:2003qh}. 

We now give results for the remaining diagrams, which
we calculate using the free-quark picture.  As before, 
we keep the external collinear quarks 
off shell by an amount $p_p^2$ when performing the matching.
We work in Feynman gauge.  With this choice of 
gauge graphs \ref{fig:1loop}(d) 
and \ref{fig:1loop}(e) vanish, 
as do the parts of \ref{fig:1loop}(b), \ref{fig:1loop}(f) 
involving soft-collinear exchange, since $n_{\pm}^2=0$.
We suppress the Dirac structure, which is always the same
as in the tree-level expression (\ref{eq:f}) after summing
over spins, and also the Wilson coefficients
$C^2(Q^2,\mu)$, which appear as a multiplicative factor.  

The non-vanishing anti-hard-collinear graphs add up to 
\bea\label{eq:Dhcbar}
&&T_{\hcbar}=-\frac{C_F\alpha_s}{4\pi}\frac{Q^2}{p_x^2}
\Bigg\{\left[-\frac{1}{\eps}-1+\ln\frac{-p_x^2}{\mu^2}\right]\nl
&&
+\left[\frac{4}{\eps^2}+\frac{4}{\eps}
\left(1-\ln\frac{-p_x^2}{\mu^2}\right)+ 2\ln^2\frac{-p_x^2}{\mu^2}
- 4\ln\frac{-p_x^2}{\mu^2}-\frac{\pi^2}{3}+8\right]\Bigg\},
\eea
the collinear graphs (including wave-function graphs)
evaluate to
\bea\label{eq:Dc}
&&T_{c}=-\frac{C_F\alpha_s}{4\pi}\frac{Q^2}{p_x^2}
\Bigg\{\left[-\frac{1}{\eps}-1+\ln\frac{-p_p^2}{\mu^2}\right] \nl
&&
+\left[\frac{4}{\eps^2}+\frac{4}{\eps}
\left(1-\ln\frac{-p_p^2}{\mu^2}\right)+ 2\ln^2\frac{-p_p^2}{\mu^2}
- 4\ln\frac{-p_p^2}{\mu^2}-\frac{\pi^2}{3}+8\right] \Bigg\},
\eea
and the soft-collinear graphs give
\be\label{eq:Dsc}
T_{sc}=
-\frac{C_F\alpha_s}{4\pi}
\frac{Q^2}{p_x^2}\left[-\frac{4}{\eps^2}+
\frac{4}{\eps}\ln\frac{p_x^2 p_p^2}{Q^2\mu^2}
-2\ln^2\frac{p_x^2 p_p^2}{Q^2\mu^2}-\pi^2\right].
\ee

The $1/\eps$ terms in the sum of all diagrams is subtracted
by the current renormalization factor $Z_J^2$ given in ({\ref{eq:ZJ}).
This is possible only after the same cancellation between logarithms in
the divergent pieces of the  anti-hard-collinear, collinear, and 
soft-collinear graphs that we observed when matching 
the SCET current.  No such cancellation occurs in the finite 
pieces, where logarithms at each scale remain.
We can interpret the finite parts as the one-loop corrections
to a convolution of functions characterizing the physics at the various scales.
This takes the form
\be\label{eq:corrections}
\frac{1}{\pi}{\rm Im} \left(T_{\hcbar}+T_{c}+T_{sc}\right)
= J^{(1)}\otimes\left[S^{(0)}\cdot f_{sc}^{(0)}\right]
 +  J^{(0)}\otimes \left[S^{(1)}\cdot f_{sc}^{(0)}
+S^{(0)}\cdot f_{sc}^{(1)}\right].
\ee 
The superscript refers to the $n$-loop correction to 
each function.  We have defined a function 
$S=D_c^2$ (see (\ref{eq:richtig})), 
which takes into account collinear effects, and 
grouped the sum of the collinear and soft-collinear 
corrections inside the square brackets.  
In this step of matching we want to obtain the one-loop
correction to the jet function $J$.  To do this rigorously,
we would first need to calculate the renormalized 
expression for the object $\left[S\cdot f_{sc}\right]$, 
using the free-quark picture.
This calculation would require a more precise formulation 
of an effective theory
defined at the soft-collinear scale.  We will not 
go through this exercise here, but rather assume
that we can construct a low-energy theory that properly
accounts for the IR physics related
to the collinear and soft-collinear fields.  The difference
between this low-energy theory and SCET is that the
anti-hard-collinear fields are absent, so the 
the matching function $J$ is given by 
imaginary part of the finite piece of $T_{\hcbar}$
in  (\ref{eq:Dhcbar}).
This imaginary part is singular at 
$p_x^2=0$ and must be interpreted in terms of distributions
to be integrated against a smooth function $F(p_x^2)$.  
To stay in the region where the SCET treatment 
is valid requires a cut on $p_x^2$, so we find it 
convenient to express the results in terms of  star 
distributions, which are defined as  \cite{DeFazio:1999sv}
\bea
\int_{\leq 0}^z dx F(x)\left(\frac{1}{x}\right)_*^{[u]}&=&
\int_0^z dx\frac{F(x)-F(0)}{x}+F(0)\ln\frac{z}{u}, \nl
\int_{\leq 0}^z dx F(x)\left(\frac{\ln(x/u)}{x}\right)_*^{[u]}&=&
\int_0^z dx\frac{F(x)-F(0)}{x}\ln\frac{z}{u}+\frac{F(0)}{2}
\ln^2\frac{z}{u}.
\eea
In terms of these distributions, we find
\bea\label{eq:Jfact}
 J^{(1)}&\otimes& \left[S^{(0)} \cdot f_{sc}^{(0)}\right]
= \frac{1}{\pi}{\rm Im} \, T_{\hcbar}\nl 
&=&Q^2 \frac{C_F\alpha_s}{4\pi}
\left[\left( 7- \pi^2\right)\delta(p_x^2)
-3 
\left(\frac{1}{p_x^2}\right)_*^{[\mu^2]}+
4\left(\frac{\ln (p_x^2/\mu^2)}{p_x^2}\right)_*^{[\mu^2]}\right].
\eea
In the limit $x\to1$ the star distribution is 
related to the plus distribution, and (\ref{eq:Jfact}) agrees  
with a corresponding expression in \cite{manohar}.  This matching
function also appears in inclusive $B$ decay in the shape-function
region  \cite{Bauer:2003pi, Bosch:2004th}.

Having obtained an expression for the one-loop jet function,
we end this section by  taking a closer look at 
the low-energy physics relevant to the
parton distribution function. If the hadronic tensor obeyed the factorization 
formula (\ref{eq:fact}), then the low-energy physics 
would depend on the collinear scale $\Lambda_{\rm QCD}^2$ 
only.  However, we have shown that the low-energy 
theory contains a product of collinear and soft-collinear
functions, what we called $\left[S\cdot f_{sc}\right]$
above, and that this object contains 
logarithms at both the collinear 
and soft-collinear scales.  The soft-collinear 
scale $\Lambda_{\rm QCD}^3/Q$  depends on the large energy $Q$,
so not all of the $Q$ dependence 
has been factorized into the hard coefficient $C(Q^2,\mu)$. 
We will explain the consequences of this in the next section.

\section{Soft-collinear effects and factorization}
\label{sec:sc}
In this section we consolidate our results
concerning the factorization of the
hadronic tensor. To summarize,  we found that for values 
of $x$ satisfying $1-x\sim \Lambda_{\rm QCD}/Q$, 
DIS involves four scales 
\be\label{eq:mscales}
Q^2 \gg Q \Lambda_{\rm QCD}\gg \Lambda_{\rm QCD}^2 \gg \Lambda_{\rm QCD}^3/Q.
\ee
The relevance of the soft-collinear
scale $\Lambda_{\rm QCD}^3/Q$ makes it impossible to derive
a factorization formula of the type  (\ref{eq:fact}).
To clarify this, we find it useful to first consider
a fictitious version of QCD, where
the collinear scale $\Lambda_{\rm QCD}^2$ is
perturbative and the soft-collinear scale is 
non-perturbative.  In this fictitious QCD, we can derive 
a factorization formula by matching onto a  
low-energy theory defined at the soft-collinear scale. 
To do so, we split up the initial-state parton momentum as 
$p_p=Q\nm/2 + p_{sc}$, where the soft-collinear residual
momentum satisfies $\np p_{sc}\sim Q(1-x)\sim \Lambda_{\rm QCD}$ 
and $\nm p_{sc}\sim M_P^2/Q\sim \Lambda_{\rm QCD}^2/Q$.
This treats the parton as a massless on-shell
collinear quark carrying momentum $Q\nm/2$, which receives
a residual momentum $p_{sc}$ through interactions with 
soft-collinear partons.
Beyond tree level the factorization formula
contains a convolution between $f_{sc}$ and 
$D_c$, in addition to that between $f_{sc}$ and $J$.
This is because $\nm p_{sc}\sim \nm p_c$, 
so the $\nm p_{sc}$ momentum can be distributed between the 
collinear and soft-collinear fields, just as the 
$\np p_{sc}\sim \np p_x$ momentum can be distributed 
between the anti-hard-collinear and soft-collinear fields.
Writing the mass scales (\ref{eq:mscales})
in terms of $Q$, $p_{sc}$, and $p_x$ we
find a factorization formula of the form 
\be\label{eq:fiction}
W\sim  H\left(\frac{Q^2}{\mu^2}\right) 
J\left(\frac{Q\np p_x -Q\np p_{sc}}{\mu^2}\right)\otimes
f_{sc}\left(\frac{\nm p_{sc}\np p_{sc}}{\mu^2}\right)\otimes
S\left(\frac{Q\nm p_{sc}}{\mu^2}\right),
\ee
where $\np p_{sc}$ and $\nm p_{sc}$ are convolution variables.
The hard function  $H$ and the soft function 
$S$ are related to the Wilson coefficients 
arising when matching the SCET current
as in (\ref{eq:richtig}),  $H=C^2$ and 
$S=D_c^2$.  The jet function $J$ is
calculated as explained in Section \ref{sec:oneloop}, 
and the soft-collinear function
$f_{sc}$ is defined by the spin-averaged matrix element 
in (\ref{eq:f}), but with $\chi_c\to \xi_{Q,sc}$. 
The jet function $J$ and the collinear function 
$S$ are not linked directly through a convolution. 
Instead, they are linked to each other only 
through a mutual convolution with the function $f_{sc}$. 
Although this formula is 
qualitatively different from  (\ref{eq:fact}),  we 
could in principle derive the 
renormalization group equations for this effective theory, 
and use them to resum all large logarithms involving
the ratio $\Lambda_{\rm QCD}/Q$. This scenario has been
mentioned in \cite{Becher:2003kh}, in analogy with 
techniques used for the off-shell Sudakov form factor 
\cite{Korchemsky:1988hd, Kuhn:1999nn}.

Our derivation of (\ref{eq:fiction}) was based on an effective
field theory approach that integrated out the larger scales 
until reaching the smallest scale, which is soft-collinear.
In real QCD it is not possible to 
use perturbation theory at the scale $\Lambda_{\rm QCD}^2$. 
This obligates us to stop the matching 
procedure at the jet scale $Q \Lambda_{\rm QCD}$ 
and lump the collinear and soft-collinear
effects into one non-perturbative function.
We have seen that some cancellations occur between 
the sum of the infinite parts of the collinear and 
soft-collinear graphs, but this does not occur in 
the finite pieces defining the matrix elements.
The hadronic tensor therefore takes the form
\be\label{eq:nonpertfact}
W\sim  H\left(\frac{Q^2}{\mu^2}\right) 
J\left(\frac{Q\np p_x-Q\np p_{sc}}{\mu^2}\right)
\otimes f\left( \frac{\Lambda_{\rm QCD}^2}{\mu^2}, 
\frac{\Lambda_{\rm QCD}^2\np p_{sc}}{Q \mu^2}\right),
\ee 
where we have inserted the physical scaling $\nm p_{sc}\sim
\Lambda_{\rm QCD}^2/Q$.  The notation makes clear that
the parton distribution function $f$ contains physics
at both the collinear and soft-collinear scales.  
We cannot match perturbatively at the scale $\Lambda_{\rm QCD}^2$,
so we have no way of deriving a low-energy theory that would
allow us to resum logarithms at the soft-collinear scale,  and 
large logarithms depending on $\Lambda_{\rm QCD}/Q$ remain.
In other words, the parton distribution function contains a 
non-perturbative dependence on the large energy $Q$. 
This is different from both (\ref{eq:fact}) and (\ref{eq:fiction}).
We conclude that a perturbative factorization of scales
is not possible in this region of phase space.

\section{Comparison with previous work}
\label{sec:comparison}

\subsection{Diagrammatic Approach}
Factorization formulas for deep inelastic scattering near
the endpoint have been derived  
using diagrammatic methods in 
\cite{Sterman:1986aj, Catani:1989ne, Korchemsky:1992xv}.  
It seems that the effective field theory calculation
leads us to different conclusions concerning the perturbative
factorization of scales.  The differences 
can be traced directly to the soft-collinear mode.  
In turn, we found that the soft-collinear
mode is relevant in a very specific region of phase space,
where $1-x$ is correlated with $\Lambda_{\rm QCD}/Q$
through the relation $1-x\sim \Lambda_{\rm QCD}/Q \sim \lambda^2$. 
To the best of our knowledge, such a power counting
has not been implemented within the diagrammatic 
approach, where one takes the 
limit $1-x\to 0$ without making the 
above-mentioned correlation.  To understand the significance
of this, recall that the effective
field theory approach led us to split the parton distribution
function into two parts according to
\be\label{eq:split}
f\to 
S\left(\frac{Q\nm p_{sc}}{\mu^2}\right)
f\left(\frac{\nm p_{sc} \np p_{sc}}{\mu^2}\right)
\sim
S\left(\frac{\Lambda_{\rm QCD}^2}{\mu^2}\right)
f\left(\frac{\Lambda_{\rm QCD}^2(1-x)}{\mu^2}\right).
\ee
A very similar observation has been made in the diagrammatic
approach, where $S$ and $f$ are related to $\phi$ and $V$ 
\cite{Sterman:1986aj}. The function $f$ is linked to the
jet function $J$ by the convolution variable $\np p_{sc}$.  
Boost invariance and dimensional
analysis require that this enter the parton 
distribution function in the combination 
$\nm p_{sc}\np p_{sc}/\mu^2$.
From this alone it is apparent that the parton distribution
function involves fluctuations at two scales, as shown 
above.  The second scale depends on $1-x$, and need not 
be soft-collinear.  One sees this clearly
from (\ref{eq:SCscalar}).  For generic values of $p_x^2\approx Q^2(1-x)$, 
the soft-collinear region is replaced by an $x$-dependent
soft region scaling as $(Q(1-x),\Lambda_{\rm QCD} \sqrt{(1-x)},
\Lambda_{\rm QCD}^2/Q$). The 
function $f$ is associated with the vacuum matrix element of 
a Wilson loop built out of
gauge fields with this scaling. 
As long as $\np p_{sc} \sim Q(1-x)\sim Q\lambda_D$,
with $\lambda_D$ numerically small but still ${\cal O}(1)$,
then both the collinear modes and these additional soft modes 
are parametrically of the order $\Lambda_{\rm QCD}^2$.  
Formulas derived with the diagrammatic approach 
are valid within this particular
large-$x$ limit.  We emphasize that this is a different 
large-$x$ limit than that considered in our work.
The non-factorizable soft-collinear effects
studied here emerge for  values of $x$ satisfying 
$1-x\sim \Lambda_{\rm QCD}/Q$. 
To apply effective field theory methods in the most straightforward way 
requires that we make this correlation, 
because only then can we calculate the results as an expansion 
in a single small parameter $\lambda.$  
This power counting for $1-x$
also ensures that we avoid the resonance region, 
where $1-x \sim \Lambda_{\rm QCD}^2/Q^2$. 
The failure of factorization for $1-x\sim \Lambda_{\rm QCD}/Q$
suggests that the most useful application
of SCET to DIS in the endpoint region might instead
use a multi-scale approach  to study
the limit $1-x\to\Lambda_{\rm QCD}/Q$ more carefully. This 
would involve replacing the soft-collinear modes by the $x$-dependent
soft modes identified above, carefully re-deriving the factorization
formula for the large-$x$ limit obtained within the diagrammatic
approach \cite{Sterman:1986aj}, 
and studying power corrections in terms of SCET operators.
This could make use of techniques similar
to those developed for the  multi-scale operator 
expansion in inclusive $B$ decay \cite{Neubert:2004dd}.

\subsection{SCET based approach}

The first application of SCET to DIS can be found in
\cite{Bauer:2002nz}, which is however limited to the standard
OPE region and has little overlap with our work. 
In \cite{manohar} Manohar carried out a SCET analysis of DIS at 
large $x$, also using a two-step matching procedure.
In the first step, the author matched
QCD in the Breit frame onto a version of SCET involving
hard-collinear fields interacting with anti-hard-collinear fields
via soft gluon exchange.  This differs from the version of SCET
used here, which involves collinear fields interacting with 
anti-hard-collinear fields via soft-collinear gluon exchange.
While our two approaches differ conceptually, our results 
for  the anomalous dimension and hard matching coefficient 
of the SCET current agree. The results are the same
because the leading-order  Lagrangians 
${\cal L}_{c+sc}, \,{\cal L}_{\hcbar + sc}$ are of the 
same form as ${\cal L}_{hc + s},\, {\cal L}_{\hcbar +s}$. 
The author used these results to derive some interesting
consequences for the anomalous dimension of the SCET current.
We disagree on some points concerning the calculation of 
the hadronic tensor in the second step of matching.
The major difference is that  \cite{manohar} found that
the effects of soft gluon exchange  
are irrelevant to the 
low-energy matrix element defining the parton distribution 
function (in the Breit frame).
Using the translation between the leading-order
Lagrangians given above, this would imply the irrelevance
of soft-collinear effects, which we did not observe here. 
This also contradicts the results for the large-$x$ limit
derived in the diagrammatic approach, where the low-energy 
matrix element splits into a product of collinear and soft
functions, often called $\phi$ and $V$ \cite{Sterman:1986aj}.

\section{Conclusions}\label{sec:conclusions}

We used soft-collinear effective theory
to examine the factorization properties of deep inelastic
scattering in the region of phase space where 
$1-x\sim \Lambda_{\rm QCD}/Q$.  An analysis of loop diagrams
in the Breit frame showed that the  
appropriate effective theory includes anti-hard-collinear,
collinear, and soft-collinear fields.  We found that 
soft-collinear effects ruin perturbative factorization. 
An attempt to use SCET to prove a perturbative factorization formula
yields instead an expression where the low-energy matrix element
defining the parton distribution function
contains a non-perturbative dependence on the large energy
$Q$.  It is therefore impossible to separate
the three scales $Q^2 \gg Q \Lambda_{\rm QCD}\gg \Lambda_{\rm QCD}^2$ in
terms of a factorization formula.  These complications 
related to the soft-collinear mode are similar 
to those found in a SCET analysis of the 
heavy-to-light form factors relevant 
to exclusive $B$ meson decay \cite{Lange:2003pk}. 
They do not appear in an analysis of factorization for 
inclusive $B$ decay in the shape-function region, 
where the presence of a heavy quark ensures that soft 
instead of soft-collinear fields are relevant to the effective
theory construction.

Our conclusions are true as long as $1-x$ is
correlated with $\Lambda_{\rm QCD}/Q$ through the relation
$1-x\sim \Lambda_{\rm QCD}/Q$. If 
$1-x$ is numerically small but still larger than $\Lambda_{\rm QCD}/Q$, 
the standard large-$x$ factorization formula derived
within the diagrammatic approach is valid.
As $1-x$ approaches the endpoint, however, non-factorizable
soft-collinear effects emerge. It would be interesting 
to use a multi-scale effective field theory approach to 
carefully re-derive the large-$x$ factorization formula using SCET,  
quantify power corrections in terms of SCET operators, and 
more carefully study the limit $1-x \to \Lambda_{\rm QCD}/Q$.

\section*{Acknowledgements}
I am grateful
to Thomas Becher, Thorsten Feldmann, Thomas Mannel, and Matthias
Neubert for useful discussions and comments on the manuscript.
This work was supported by the DFG Sonderforschungsbereich SFB/TR09 
``Computational Theoretical Particle Physics''.

\end{document}